# A unified method of data assimilation and turbulence modeling for separated flows at high Reynolds numbers

## Zhiyuan Wang[1,2,3], Weiwei Zhang[1,3]

1 School of Aeronautics, Northwestern Polytechnical University, Xi'an 710072, China

2 Beihang Univ Aeronaut & Astronaut, Fluid Mech Key Lab Educ Minist, Beijing 100191, Peoples R China

3 National Key Laboratory of Aerodynamic Design and Research, Northwestern Polytechnical University, Xi'an 710072, China

**Abstract:** In recent years, machine learning methods represented by deep neural networks (DNN) have been a new paradigm of turbulence modeling. However, in the scenario of high Reynolds numbers, there are still some bottlenecks, including the lack of high-fidelity data and the convergence and stability problem in the coupling process of turbulence models and the RANS solvers. In this paper, we propose an improved ensemble kalman inversion method as a unified approach of data assimilation and turbulence modeling for separated flows at high Reynolds numbers. The trainable parameters of the DNN are optimized according to the given experimental surface pressure coefficients in the framework of mutual coupling between the RANS equations and DNN eddy-viscosity models. In this way, data assimilation and model training are combined into one step to get the high-fidelity turbulence models agree well with experiments efficiently. The effectiveness of the method is verified by cases of separated flows around airfoils(S809) at high Reynolds numbers. The results show that through joint assimilation of vary few experimental states, we can get turbulence models generalizing well to both attached and separated flows at different angles of attack. The errors of lift coefficients at high angles of attack are significantly reduced by more than three times compared with the traditional SA model. The models obtained also perform well in stability and robustness.

**Keywords**: turbulence modeling; machine learning; data assimilation; ensemble kalman inversion;

# 1. Introduction

Computational fluid dynamics（CFD） has made great progress in the past decades, and plays an increasingly important role in scientific research and engineering in aeronautics and many other fields. Most of the practical applications require the accurate characterization of the dynamics of turbulence. Despite advancements in computing resources, large eddy simulations (LES) and direct numerical simulations (DNS) are still decades away from being applicable in high Reynolds number flows. Reynolds-Averaged Navier-Stokes (RANS) simulation is still the dominant tool for industrial problems[1]. Since the development of effective and applicable turbulence models has become a critical issue. Traditional RANS models, especially eddy viscosity models like SA model[2], $k-\varepsilon$ / $k-w$ model[3-5] and Menter's SST model[6] have been widely used in engineering applications and are proven to be quite effective. However, it is well-recognized that complex effects such as flow separation, secondary flows, etc. are poorly modeled.

The discrepancy of RANS models comes from the empiricism and assumptions that are violated in many practical applications, which leads to the inaccurately prediction of the distributions of eddy viscosity or Reynolds stress tensors. In order to overcome the defects of traditional RANS models, machine learning methods have been introduced for turbulence modeling in recent years[7]. Researchers have developed several methods based on DNS or LES solutions that can significantly reduce the model discrepancy. Ling and Templeton[8] used DNS and LES solutions to identify where RANS turbulence models have large errors. Wang et al[9-11] proposed a physics-informed machine learning approach for correcting the Reynolds stress tensor based on DNS solutions. The resulting predictions were shown to be more accurate than the traditional RANS model even for different geometries. Ling et al[12,13] proposed a Tensor Basis Neural Networks (TBNN) model to learn the Reynolds stress anisotropy tensor from DNS solutions. The TBNN model embedded Galilean invariance into the network and showed more accurate predictions than the traditional RANS model. Although the above machine learning approaches have shown exciting progress in turbulence modeling, the vast majority of them depend on detailed LES or DNS training data, which is typically unavailable for even moderate Reynolds numbers. Zhu et al[14] constructed a pure data-driven turbulence model based on RANS solutions at high Reynolds numbers to replace the traditional RANS models. The proposed model has a certain generalization ability for the computational state and geometries with higher computational efficiency than SA model. But due to the limitations of the training data, the accuracy of the results cannot exceed that of traditional RANS models. To sum up, the acquisition of high-fidelity flow field data is a key bottleneck of turbulence modeling at high Reynolds numbers.

In engineering turbulence problems at high Reynolds numbers, data generated from physical experiments becomes an important reference for data-driven turbulent modeling. However, the fine measurement of all physical quantities in the boundary layer is still very challenging. Data generated from experiments is usually noisy and very limited. Given above problems, several data assimilation/field inversion methods are proposed to reconstruct high-fidelity flow field data from limited and noisy experimental data. The data assimilation methods can be mainly divided into adjoint Methods[15] and Bayesian inversion[16,17]. Duraisamy et al.[18-20] proposed a field inversion and machine learning (FIML) method. The approach consists of two main steps, inferring the spatial distribution of model discrepancy $\beta(x)$ as a multiplier of the production term in SA equation using a discrete adjoint approach and then training the machine learning model based on the inferring results. Continuous adjoint method is also used to infer $\beta(x)$ [21], which is considered to be more effective than discrete adjoint approach. Foures et al.[22] proposed a data assimilation method based on the variational formulation and the Lagrange multipliers approach to reconstruct the mean flow field aiming at minimizing the error between the DNS data and the numerical solution of RANS equations. Besides adjoint methods, Bayesian inversion approach is also used for data assimilation[23]. A variation of Bayesian inversion is ensemble kalman filter. Kato et al.[24] utilized ensemble transform kalman filter (ETKF) to correct the angles of attack (AOA) and the Mach number as well as the eddy viscosity distributions according to experimental data. The prediction accuracy is significantly improved. In addition, Kato and Obayashi[25] also applied ensemble Kalman filter method to correct the coefficients of traditional RANS models. Deng et al.[26] correct the coefficients of several RANS models by ensemble kalman filter. The prediction accuracy of different RANS models after correction is consistent with the high-fidelity data. Xiao et al.[27,28] proposed a regularized ensemble Kalman method to add some additional constraints to the cost function such as Spatial smoothness and prior values. The results are improved compared with the original ensemble Kalman method. In addition, some other methods have also been used for field inversion. Such as the POD-inverse method proposed by Liu et al.[29] and the D-dark method proposed by Li et al[30].

When applying data assimilation methods above for turbulence modeling, data assimilation and model training are two separate steps. We need to solve the inverse problem to infer the full-field high-fidelity data first. Then the machine learning model is trained on the frozen datasets. Such two-step approach is effective for solving the data acquisition issue at high Reynolds numbers. However, the inconsistency caused by the separation of inference and modeling training[31] brings several problems: 1) the inferred solutions are dependent of computational grids and particular flow solution, this may lead to unphysical issues like non-smoothness and negative values of eddy viscosity as well as conflict results of different flow solutions. 2) There are no guarantees that the inferred data is learnable for a specific model configuration. 3) Small errors of model output may be dramatically amplified during the coupling process of RANS solver and turbulence models. Such stability problem is also is referred to as the ill-conditioning of the RANS operator[32,33] in the literature. Furthermore, the convergence and stability issues in the coupling process of turbulence models and the RANS solvers is particularly prominent in separated flows at high Reynolds numbers[30]. the unphysical and unlearnable problems will further deteriorate the convergence and stability issues.

In order to deal with the above drawbacks of the two-step approach. RANS solvers are involved in the model training process. In this way, indirect data such as force coefficients is used to find the models perform best when embedded in the RANS solvers. Such a strategy is referred to as "model-consistent learning"[31].

In the process of model-consistent learning, RANS equations are solved in every iteration to get the numerical solutions of the current turbulence model, which is then used to evaluate the model discrepancy and optimize the model parameters. In this way, data accessible in experiments including velocity and force coefficients can be used directly to train the data-driven turbulence model. To solve such an optimization problem, gradient descent method is usually used. The neural networks-contributed gradient can be easily obtained by the backpropagation algorithm. While the calculation of the RANS solver-contributed gradient is relatively more difficult. Currently, discrete adjoint method[34,35], continuous adjoint method[36] and ensemble approximation method[37] are both proposed to obtain the sensitivities of RANS equations. Then the full gradient is calculated by the chain rule and is used for gradient-based optimization. Adjoint methods including discrete and continuous adjoint method are effective but intrusive, requires significant effort for different solvers. Ensemble approximation method is non- intrusive in the price of lower accuracy. In addition, there are some other approaches for the same purpose of model-consistent learning have been proposed for turbulence models represented as symbolic expressions[38].

Recently, ensemble kalman inversion approach is proved to be effective for model-consistent learning[39,40]. Zhang et al.[40] applied ensemble kalman method for turbulence modeling based on DNS solutions and demonstrated the effectiveness and advantages of the method. On the one hand, the forward simulation of a group of ensemble members are conducted in every iteration instead of the computation of gradient. Ensemble kalman method is non-intrusive, which means it can be implemented straightforwardly to different solvers without any additional modification of the codes. On the other hand. As a gradient approximation approach, ensemble kalman method also implicitly use the second-order gradient for optimization, which leads to better performance than adjoint methods.

However, there are still several issues applying ensemble kalman approaches for turbulence modeling, especially at high Reynolds numbers. The optimization capability and numerical stability of ensemble Kalman approaches depends significantly on the initial ensemble members. What's more, in the context of turbulence modeling, the generation method of initial NN model members also significantly influence the physical properties of the turbulence models as well as the convergence and stability performance, which will be much more prominent in separated flows at high Reynolds numbers. Model members with unsatisfactory parameters may result in unphysical outputs as well as poor convergence and stability performance. Additionally, in implementations, ensemble kalman approach requires a sufficient amount of ensemble members so that the effectiveness and numerical stability can be guaranteed. Every model member needs to be coupled with RANS solver until the given convergence tolerance is reached in each iteration until the training procedure convergent. High computational cost in tens to hundreds of iterations the significantly limits the efficiency of the method.

To sum up, at low Reynolds numbers, DNS solutions are usually accessible for data-driven turbulence modeling. While in separated flows at high Reynolds numbers, there are still several challenging bottlenecks and limitations of current methods. In this paper, we proposed an improved ensemble kalman inversion method to realize the unification of data assimilation and model training procedure. The trainable parameters of DNN eddy-viscosity model are optimized according to the given experimental surface pressure coefficients in the framework of mutual coupling between the RANS solvers and turbulence models. It helps to significantly improve the computational efficiency as well as alleviate the prominent issues such as the generalization and stability performance of the model at high Reynolds numbers.

The rest of this paper is organized as follows. The architecture of the DNN eddy-viscosity model and the improved ensemble kalman inversion method are presented in Section 2. The cases for testing the performance of the proposed unified method of is detailed in Section 3. The computational efficiency of the method and the generalization capability of the models are discussed in Section 5. Finally, conclusions are provided in Section 6.

## 2. Methodology

### 2.1 the overall introduction

In our perspective, the method proposed in this paper is a unified approach of data assimilation and model training. Fig.1 is the overall Schematic diagram. In the mutual coupling framework of RANS solver and DNN eddy viscosity models, data assimilation algorithms such as the improved ensemble kalman inversion method proposed in this work is introduced to optimize the DNN trainable parameters according to the high-fidelity experimental data. In this way, the DNN model in the best agreement with the experiments is obtained directly through data assimilation methods. In this chapter, we will introduce the detailed methods including the framework of our DNN eddy viscosity model, the coupling mode between DNN models and RANS solvers, as well as the improved ensemble kalman inversion method proposed in this paper.

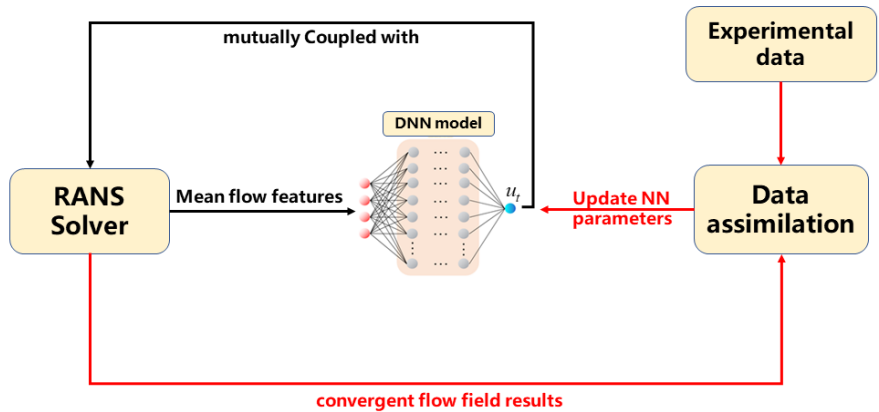

**Fig.1.** the overall schematic diagram of the method

### 2.2 The framework of the DNN eddy viscosity model

The turbulence modeling framework refers to our previous works[14,41-43]. Full connected deep neural network also known as Multilayer Perceptron (MLP) is chosen to construct the data-driven eddy viscosity model. The schematic diagram of full connected deep neural network is shown in Figure 2.

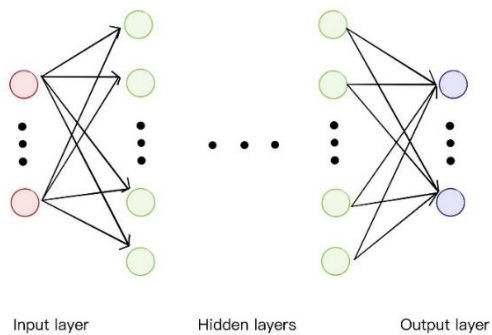

**Fig. 2.** Schematic diagram of fully connected neural network

**Input and output features**

Machine learning algorithms rely heavily on the input features. Well selected features can speed up the training process, improve the accuracy and the generalization ability of the model. Input features with clear physical meaning are especially critical for the performance of data-driven turbulence models. In this work, we refer to our previous work of Sun et al.[41] . Eight input features are constructed and selected out of the consideration of physical properties and performance of the model. The input features are shown in Table 1.

**Table 1** The input features of the model, where U and V are the velocity in the x,y direction, P is the pressure, ρ is the density, γ=1.4 is a constant, Y is the y coordinate, *dis* is the minimum distance from the grid node to the wall, and *sig* is the symbol Function, *tanh* is the hyperbolic tangent function

| # | Formula | Description |
|---|---|---|
| $q_1$ | $U$ | X component of velocity |
| $q_2$ | $\sqrt{(\frac{\partial U}{\partial y} - \frac{\partial V}{\partial x})^2}$ | Norm of vorticity |
| $q_3$ | $\frac{\gamma P}{\rho^\gamma} - 1$ [44] | Entropy |
| $q_4$ | $\arctan(\frac{V sig(Y)}{U})$ [44] | Direction of velocity |
| $q_5$ | $\sqrt{2(\frac{\partial U}{\partial x})^2 + 2(\frac{\partial V}{\partial y})^2 + (\frac{\partial U}{\partial y} + \frac{\partial V}{\partial x})^2}$ | Strain rate |
| $q_6$ | $dis^2 q_2 (1 - \tanh(dis))$ | Empirical function of dis |
| $q_7$ | $sig(Y) * [-V + U * \tan(\alpha)]$ [44] | Velocity projection |
| $q_8$ | $e^{\sqrt{\frac{Dref_1}{\min(dis)}}} * \sqrt{\frac{Dref_0}{Dref_2}} - 2$ | Empirical function, where $Dref_0 = \frac{1}{\sqrt{Re}}$ $Dref_1 = \min(dis, Dref_0)$ $Dref_2 = \max(dis, Dref_0)$ |

The output feature is also built to highlight the importance of the eddy viscosity in the boundary layer and reduce the discrepancy of the data distribution under different Reynolds numbers. To this end, the following transformation have been taken:

$$\hat{\mu}_t = \mu_t \cdot e^{\sqrt{Re^{-1/2} dis}} \qquad (1)$$

The values of eddy viscosity in the near-wall surface are amplified after the transformation, thus reflecting the higher weight of the near-wall region.

Then, the input and output data are both normalized to eliminate the difference in magnitude between features and improve modeling efficiency:

$$\hat{x} = \frac{x - x_{min}}{x_{max} - x_{min}} \tag{2}$$

where $x$ denotes a feature, $x_{max}$ denotes the maximum value of the feature, and $x_{min}$ denotes the minimum value of the feature. $\hat{x}$ is the normalized feature.

**Training method**

Gradient decent methods based on backpropagation algorithm like the Adam optimizer are widely used to train neural networks. In this work, the Adam optimizer is used in the pre-training and fine-tune steps to obtain a group of ensemble members of DNN models, which will be introduced in detail in 2.3.

When using neural networks for regression tasks, the commonly used loss functions are mean absolute error (MAE) or L1 loss, mean square error (MSE) or L2 loss as follows:

$$L_1(y, \hat{y}) = \frac{1}{N} \sum_{i=1}^{N} |y_i - \hat{y}_i| \tag{3}$$

$$L_2(y, \hat{y}) = \frac{1}{N} \sum_{i=1}^{N} (y - \hat{y})^2 \tag{4}$$

where $y$ is the true value and $\hat{y}$ is the predicted value.

In turbulence modeling problems, some other constraints need to be considered. Here, except of the MSE error, two more terms are added in the loss function as follows:

$$Loss(y, \hat{y}) = \frac{1}{n} \sum_{i=1}^{n} (y - \hat{y})^2 + \frac{1}{n} \sum_{i=1}^{n} \max(0 - \hat{y}, 0) + \frac{1}{n} \sum_{i=1}^{n} \max(\hat{y} - 1, 0) \tag{1}$$

The second term represents the loss arising from the fraction of model predictions less than 0. a priori knowledge is that the eddy viscosity will not be a negative value. Therefore, it is necessary to put some constraints on the negative values in the model predictions. When the prediction is negative, the second term generates losses to guide the model to learn in a more physically consistent direction. The third term has a similar meaning to the second term and serves to constrain the model to produce predictions greater than one. This is mainly since we normalize the data by deflating the training data to between [0,1].

**Coupling mode between DNN models and RANS solvers**

After the modeling process, the DNN eddy viscosity model is embedded in the RANS solver and calculates the eddy viscosity according to the mean flow variables in each iteration to completely replace the traditional turbulence model. The coupling computation is not over until the given convergence tolerance is reached.

The RANS calculations are conducted with our in-house code which is based on the cell-centered finite volume method. In this paper, Roe scheme[45] is used to evaluate the convective fluxes, and the interface values are reconstructed by the second-order least-squares approach. The viscous fluxes are simply discretized with the standard central scheme. For the semi-discretized form of the governing equations, the implicit Symmetric Gauss-Seidel scheme[46] is adopted to march in time integral, and the local time stepping and residual smoothing techniques are also employed to accelerate convergence. Our in-house RANS solver has been validated completely by simulating many complex configurations in engineering at high Reynolds numbers, and the details can refer to our previous work[43,47,48].

Furthermore, several tricks are used in the coupling process to improve the convergence and stability performance. For example, we use the spatially smoothing method to take the weighted average of the vortex coefficients on the current grid and all its neighboring grids as the final vortex coefficient value on the current grid. This method can make the distribution of vortex viscosity on the space smoother and reduce the generation of singular values.

## 2.3 Ensemble kalman inversion method

Data assimilation methods are introduced to turbulence problems for the solution of the inverse problem of inferring the flow field variables such as eddy viscosity or Reynolds stress given high-fidelity experimental data. This corresponds to the inference step in the two-step approaches for turbulence modeling. For the unified approaches, the inverse problem turns into the inference of DNN trainable parameters that produces outputs in the best agreement with the experimental data when embedded in RANS solvers.

In this work, such inverse problem is solved based on ensemble kalman inversion method (EKI). The principle of the method can be traced back to ensemble kalman filter (EnKF). EnKF is first proposed by Evensen[17] in 1994. In data assimilation community, EnKF has been one of the most popular methods for its robustness, ease of implementation and numerical evidence of its accuracy in nonlinear problems. EnKF and several its variants have been widely used in atmosphere, geoscience and hydrological data assimilation[49-53]. The application of an iterative ensemble Kalman method for the solution of a wide class of inverse problems was proposed by Iglesias et al.[54] in 2013. The iterative ensemble Kalman method is referred as ensemble kalman inversion (EKI). Several variants of EKI were proposed subsequently to improve the capability[55,56]. Kovachki et al.[57] proposed a EKI approach as a derivative-free technique for machine learning tasks. Applications of the approach include offline and online supervised learning with deep neural networks, as well as graph-based semi-supervised learning. In this work, we refer to the EKI method as the optimizer.

## 2.3.1 The classical EKI method

The classical EKI method is introduced first. The inverse problem can be described as the finding of the $u$ given obversions in the form of

$$y = \varsigma(u) + \eta \qquad (2)$$

Where $\varsigma: X \to Y$ is the forward response operator mapping the unknown $u$ to the response/observation space. For example, the forward response that arises from physical systems described by the solution of a PDE system. X and Y are Hilbert spaces. $\eta \in Y$ is a noise and $y \in Y$ the observed data. $\eta$ is assumed a realization of a mean zero random variable whose covariance $\Gamma$ is known.

Artificial dynamics based on state augmentation are constructed to solve the inverse problem described above. The space $Z: X \times Y$ and the mapping $\Xi: Z \to Z$ are defined:

$$\Xi(z) = \begin{pmatrix} u \\ \varsigma(u) \end{pmatrix} \text{ for } z = \begin{pmatrix} u \\ p \end{pmatrix} \in Z \qquad (3)$$

The artificial dynamics are constructed as follows:

$$z_{k+1} = \Xi(z_k) \qquad (4)$$

The observed data is assumed to has the form:

$$y_{k+1} = Hz_{k+1} + \eta_{k+1} \qquad (5)$$

Where the projection operator $H: Z \to Y$ is defined by $H = (0, I)$ and $\{\eta_n\}_{n \in \mathbb{Z}^+}$ is an i.i.d sequence with $\eta_1 \sim N(0, \Gamma)$.

The iterative ensemble kalman algorithm is as follows:

**Algorithm 1**: Iterative ensemble method for inverse problems.

Initialize the ensemble: $\{z_0^{(j)}\}_{j=1}^J$

for k = 1

(1) Prediction step.

Propagate, under the artificial dynamics

$$z_{k+1}^{(j)} = \Xi(z_k^{(j)}) \qquad (10)$$

Calculate the sample mean and covariance of the ensemble:

$$\bar{z}_{k+1} = \frac{1}{J} \sum_{j=1}^{J} \hat{z}_{k+1}^{(j)} \qquad (11)$$

$$C_{k+1} = \frac{1}{J}\sum_{j=1}^{J} \hat{z}_{k+1}^{(j)}(\hat{z}_{k+1}^{(j)})^T - \bar{z}_{k+1}\bar{z}_{k+1}^T \tag{12}$$

(2) Analysis step.

The kalman gain is obtained:

$$K_{k+1} = C_{k+1}H^*(HC_{k+1}H^* + \Gamma)^{-1} \tag{13}$$

Where $H^*$ is the adjoint operator of $H$.

The observed data is then introduced to update each ensemble member:

$$\begin{aligned}z_{k+1}^{(j)} &= I\hat{z}_{k+1}^{(j)} + K_{k+1}(y_{k+1}^{(j)} - H\hat{z}_{k+1}^{(j)}) \\ &= (I - K_{k+1}H)\hat{z}_{k+1}^{(j)} + K_{k+1}y_{k+1}^{(j)}\end{aligned} \tag{14}$$

Compute the mean of the parameter update:

$$u_{k+1} = \frac{1}{J}\sum_{j=1}^{J} H^{\perp}z_{k+1}^{(j)} = \frac{1}{J}\sum_{j=1}^{J} u_{k+1}^{(j)} \tag{15}$$

and check for convergence.

In the application to deep neural networks in this work, $u = \begin{pmatrix} w \\ b \end{pmatrix}$ where $w$ and $b$ are the weights and bias. $\varsigma(u)$ represents the propagation in RANS solvers. y represents the experimental data. in this work, y is the pressure coefficients on the airfoil surface.

**2.3.2 ensemble Initialization method based on transfer learning**

In the implementation of ensemble kalman method, a group of ensemble members is necessary for the estimation purpose. The creation of the Initial ensemble members requires perturbing every parameter to be estimated. The initial ensemble members are critical for the performance and numerical stability of the method. In existing methods of data assimilation and inversion approaches based on ensemble kalman method, randomly sampling is used to perturb the state parameters[26,37]. Additionally, parameters can also be perturbed indirectly, but also in a randomly sampling way. Such as perturbing the eddy viscosity distribution by randomly sampling the karman constant of the traditional SA model[24]. In the application of ensemble kalman inversion method to train neural networks, randomly sampling is also used[57].

However, as mentioned in section 1, the initial ensemble members are particularly critical for turbulence modeling at high Reynolds numbers. On the one hand, as a physical problem, the randomly sampling of thousands of parameters of DNN has no guarantees of the physical properties of the turbulence model. The generalization capability of the turbulence models with randomly sampling parameters may also be very poor.

On the other hand, whether in the offline model-consistent training process or the online prediction process, the DNN turbulence model need to be embedded in RANS solvers. The convergence and stability issues are prominent especially in separated flows at high Reynolds numbers. There are also no guarantees that DNN models with randomly sampling parameters have good performance of convergence and stability when embedded in RANS solvers.

In this work, we proposed a novel perturbation method based on transfer learning. Transfer learning[58] aims to apply knowledge or patterns learned in one task to different but related tasks. In this work, we use transfer learning[58] as a perturbation method. Specifically, the pre-training and fine-tune method[59] is used on different datasets. The Schematic of the ensemble initialization method is shown in Fig. 7.

First, in the pre-training step, a proper pre-training dataset $D_0$ of RANS solutions with traditional SA model is constructed. Convergent flow fields at different angle of attack and a few of flow fields that have not reached the convergence tolerance is used in this work out of the purpose of improving the convergence and generalization performance of the model. DNN eddy viscosity model is trained on the pre-training dataset by Adam optimizer. We name it the base model. Then, in the fine-tune step, the base model is transferred to different datasets $\{D_1, D_2 \cdots D_j\}$ to perturb the parameters. In this paper, we construct the $\{D_1, D_2 \cdots D_j\}$ by RANS simulation with SA model at several different angle of attack respectively. In the fine-tune step, the optimizer, loss function and normalization method are all the same as before. By constructing the loss function as described in the second part of this section, physical constrains can be easily imposed to every ensemble member. We can get different ensemble members by retraining on different dataset as well as retraining on the same dataset but with different epochs, which corresponds to different perturbation directions and magnitudes of the DNN. Although the parameters created in this way may don't follow a Gaussian distribution. The effectiveness of ensemble kalman method in non-Gaussian problems has been widely validated[60]. The Schematic of the ensemble initialization method is shown in Fig. 7.

Furthermore, to get diverse ensemble members, the fine-tune datasets can also be constructed through several ways as follows: 1) RANS solutions at different conditions include different inflow conditions and boundary conditions. 2）RANS solutions with a specific turbulence model but with different coefficients of the model. 3）RANS solutions with different turbulence models.

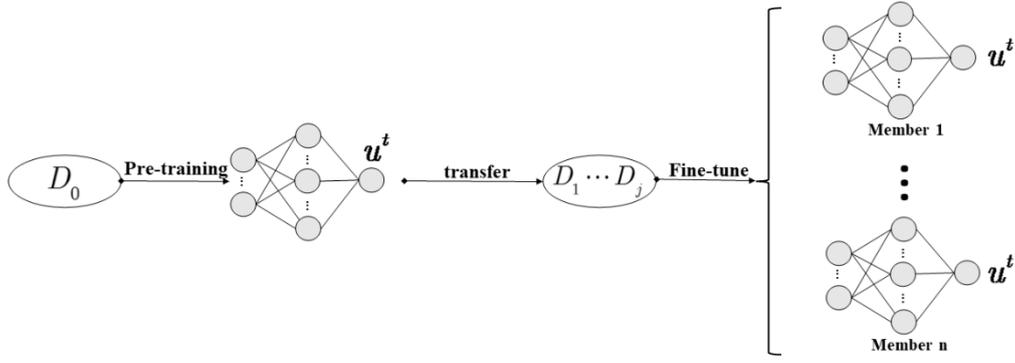

**Fig 3.** Schematic of the ensemble initialization method based on transfer learning

Ensemble members created by transfer learning method described above has several significant advantages compared with randomly sampling method. First, the pre-training and the fine-tune step are all based on RANS solutions in the well-designed modeling framework. The information contained in the RANS solutions is learned as a priori knowledge. And the difference of different RANS solutions is used to create diverse members. Also, physical constraints are easily imposed by the design of the loss function. So, the physical property and the convergence and stability performance of the model can be guaranteed to some extent. A group of diverse DNN models with good prior values are provided for the solution of the inverse problem in this way. This is of great significance to improve the performance and efficiency of the method and guarantee the performance of the DNN turbulence model.

### 2.3.3 A regularized method by the setting of observation covariance

In EKI method, the following minimization problem is solved in a derivative-free way[57]:

$$J(x) = \|u - u^f\|_{P^{-1}}^2 + \|y - \varsigma(u)\|_{\Gamma^{-1}}^2 \quad (16)$$

Where $u^f$ is the prior value of $u$, $P$ and $\Gamma$ the covariance matrices of the state $u$ and the observation y.

When solve an inverse problem using EKI method, $\Gamma$ is an important parameter for the numerical stability and convergence performance of the optimizer[55,56,61] instead of just a measure of uncertainty in traditional data assimilation methods. It is analogous to the hyperparameters in the training process of neural networks such as the step size. The value of $\Gamma$ can be set to vary with the number of iteration in previous works. In this work, we proposed a setting strategy of $\Gamma$. The value of $\Gamma$ varies not only with different iterations but also with different datasets and data points at different spatial positions. The strategy is based on the prior knowledge of the model and observation data. The setting of $\Gamma$ can be considered as an approach to give different weights to different data in the inference process. We can set smaller values of some data points in the covariance matrix $\Gamma$, which means the data is considered relatively more accurate and since is given larger weight compared with observation data with larger values in $\Gamma$.

As a physical problem, different property of the observation data in different flow conditions such as different AOA or different positions in the flow field are taken into account. For example, we know that the errors of RANS solutions mainly reflected in separated flows. So, in the inference process, we can give larger weights to the observation data in separated flows by the setting strategy. Also, when we make use of the experimental data in different AOA to train a model jointly, the setting strategy can be used for tradeoff to improve the efficiency and numerical stability.

To sum up, transfer learning and the setting strategy of the observation covariance are introduced to improve the traditional ensemble kalman inversion method. The proposed method for training the DNN eddy viscosity model with experimental observation data is schematically illustrated in Figure 4. The improved ensemble Kalman inversion method proposed in this paper consists of the following steps:

(a0) Pre-training step：Train the base model on a dataset $D_0$ which is generated from RANS simulations with traditional turbulence models at different flow states.

(a1) Fine-tune step：Retrain the base model on diverse datasets to obtain a group of perturbed parameters.

(b) coupling step：Every ensemble member is embedded in RANS solvers to obtain the prediction corresponding to high fidelity observation data Y.

(c) Updating step: Model parameters are updated Through statistical analysis of the predicted quantities and the comparison to observation data Y.

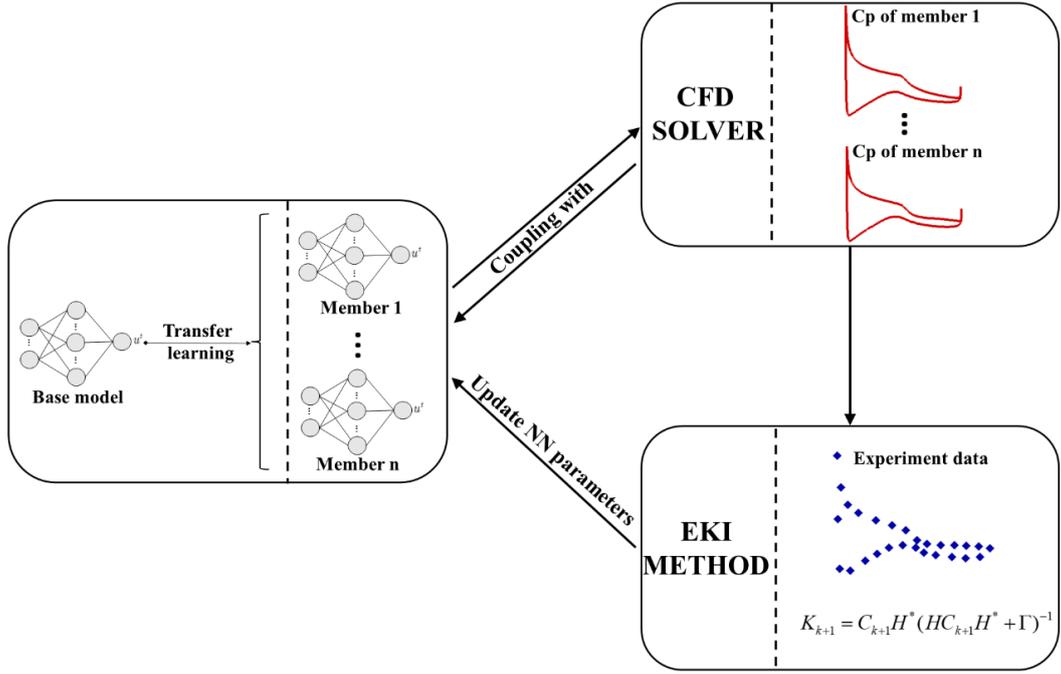

**Fig.4**. Schematic of the improved ensemble kalman inversion method

Furthermore, from a machine learning perspective, the method proposed in this paper can be considered as a dual transfer learning method. Firstly, transfer learning method is used in the ensemble initialization step. The base model trained on the pre-training dataset is transferred to different datasets of RANS solutions to obtain a group of ensemble members. Secondly, the ensemble members are transferred to experimental dataset to infer the high-fidelity DNN model. Such a procedure can also be classified as transfer learning with ensemble kalman inversion as the training method instead of gradient decent methods. Such a dual transfer learning strategy makes comprehensive use of the following three aspects of information: 1) the prior knowledge of RANS solutions simulated with traditional turbulence models. 2) the difference between flow fields generated in different flow conditions. 3) physical information implied in high-fidelity but sparse experimental data. The combination of the three above aspects can also be considered as a data fusion approach.

## 3. Numerical cases

In this section, the proposed method is applied to the S809 airfoil at high Reynolds numbers. First, the method is validated with observation data of a single angle of attack. Then more states are taken into account.

The computational mesh is shown in Fig. 9, in which the total number of elements is 36,077, with 400 nodes on the airfoil surface. There are 40 layers of mesh in the boundary layer with the growth rate 1.1, and the first grid height is $8 \times 10^{-6}$. The free stream Mach number $Ma = 0.2$, and the Reynolds number $\mathrm{Re} = 2 \times 10^6$.

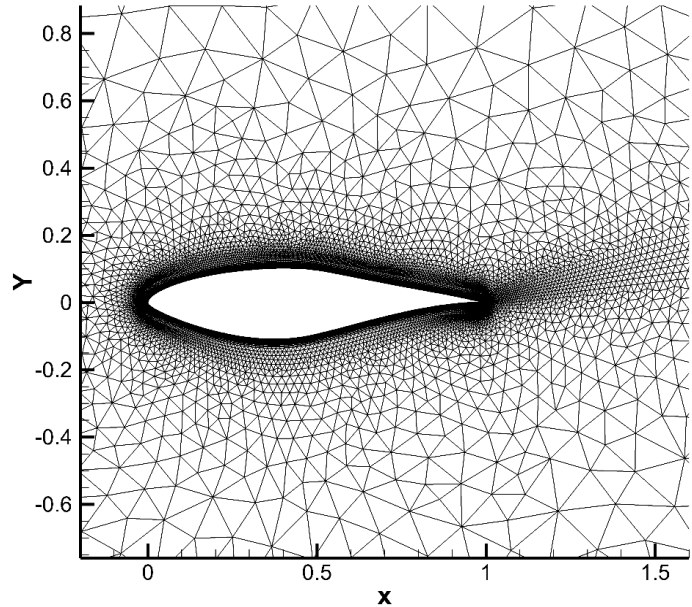

**Fig. 5**. The computational mesh for S809 airfoil.

As mentioned before, accurate modeling and prediction of separated flows remains an outstanding issue for traditional turbulence models. At high angles of attack, a strong adverse pressure develops on the upper surface of the S809 airfoil. We compute the turbulent flow fields of the S809 airfoil with the traditional SA model at angles of attack from 0 to 14 degrees. The comparisons of lift coefficients with different angles of attack between the SA model and experimental data are shown in Fig. 6. For the traditional SA model, the computed lift coefficients are consistent with the experimental data in the condition of attached flows at the angles of attack of $\alpha \leq 9°$. However, when the angle of attack is larger than 9°, the results of SA model are quite different from the experimental data. This is mainly because the SA model has difficulty predicting the separation point and the size of the recirculating region.

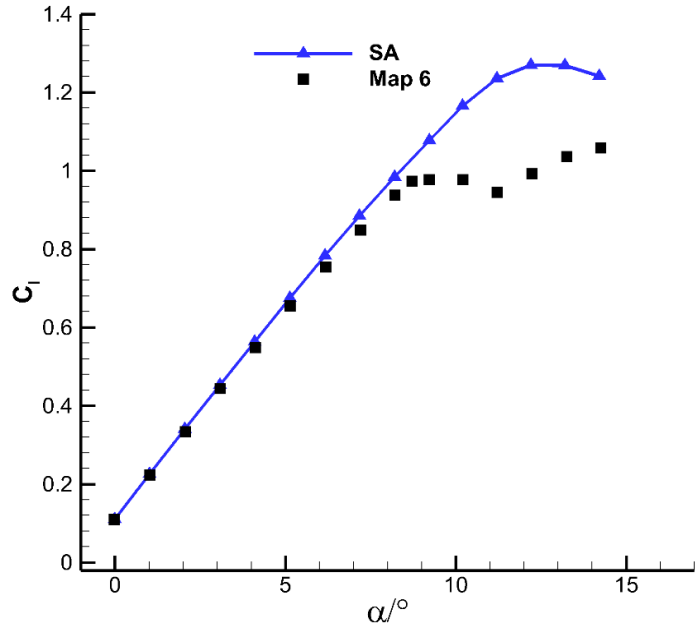

Fig. 6. The comparisons of lift coefficients with the angles of attack between SA model and experimental data.

## 3.1 Modeling and assimilation on a single AOA

The proposed method is first applied with the high-fidelity data of a single AOA. The experimental pressure coefficients at 10.20° are used here. Following the method described in section 2, we first train a base model on a pre-training dataset of RANS solutions, which is obtained at different angles of attack. And then the base model is retrained on different fine-tune datasets. Each of the fine-tune dataset contains the RANS solution of a single AOA. The base model is retained on each dataset for 2-20 epochs to perturb the parameters with different magnitude. In this way, 60 initial ensemble members with satisfactory physical properties and convergence performance are obtained. The ensemble members are then embedded in RANS solvers respectively in the same flow condition of the experimental data to get the flow fields. The prediction of every ensemble member is then compared with the experimental data at 10.20° to update the parameters of the DNN model using ensemble kalman inversion method.

With the proposed method, only a single iteration is enough for training the turbulence model on experimental data in this case. The computational cost is significantly reduced compared with traditional ensemble kalman methods which usually need at least tens of iterations. This will be discussed later in section 4.

Fig. 7 shows the comparisons of pressure coefficients between the two models and the experimental data. As stated before, the $C_p$ simulated by SA model is quite inaccurate on the upper surface. The size of the recirculating region is grossly underestimated. In contrast, the prediction results of the DNN model obtained are consistent with the experimental data. And the recirculating region is accurately predicted, which can be found from the comparison of velocity distributions and streamlines for the DNN model and the SA model in Fig.8.

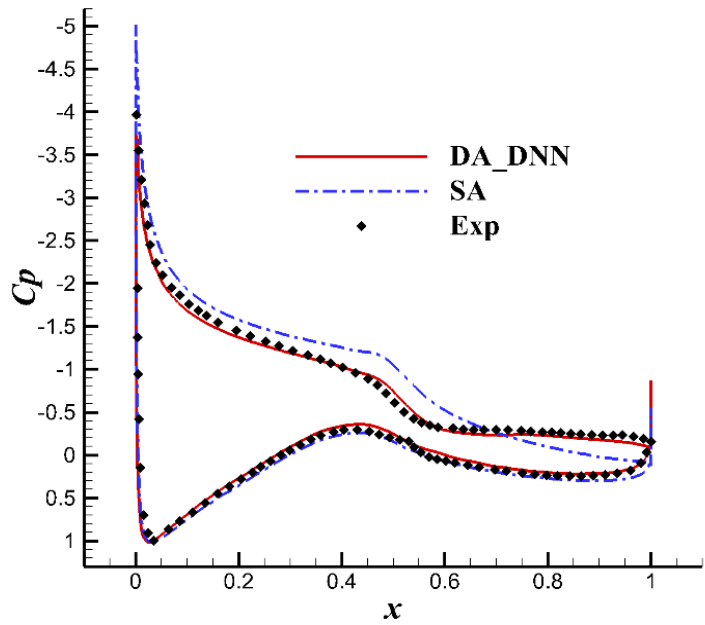

**Fig.7.** The comparisons of the pressure coefficients between experimental data and numerical results at 10.20°．

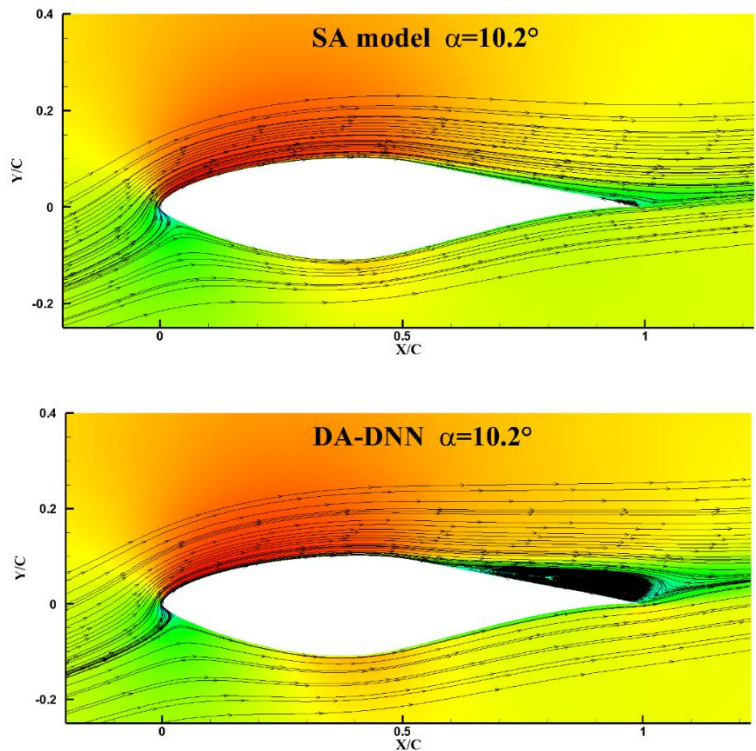

**Fig.8.** The comparisons of velocity distributions and streamlines between the DNN model and the SA model at 10.20°

The effectiveness of the method is verified through the Modeling and assimilation on a single AOA. However, the generalization capability of the model trained on experimental data of a single state is very limited. For example, the DNN model we get using the experimental pressure coefficient at 10.2° generalizes well to separated flows at higher AOA but fails to predict the attached flows at lower AOA. Since it's necessary to introduce high-fidelity data of more states.

## 3.2 Joint Modeling and assimilation on multiple states

Adjoint method has been used to train the model with high-fidelity data of multiple states[62]. In this work, a more extreme training strategy compared with that in the literature is taken. Although there are no methodological difficulty adding more high-fidelity data, only the experimental pressure coefficients at the angle of attack of 9.22° and 10.20° are used to test the proposed method. That means the simulations of other angles of attacks are all extrapolation for the DNN model.

The initial ensemble members are the same as the first case but are embedded in RANS solvers to obtain the solution at 9.22° and 10.20° respectively. In this case, a single iteration is also enough. The results of the pressure coefficients at 9.22° and 10.2° are shown in Fig.9. We can see from the results that through joint modeling and assimilation, the results given by the DNN model are in consistent with the experimental data at both training states.

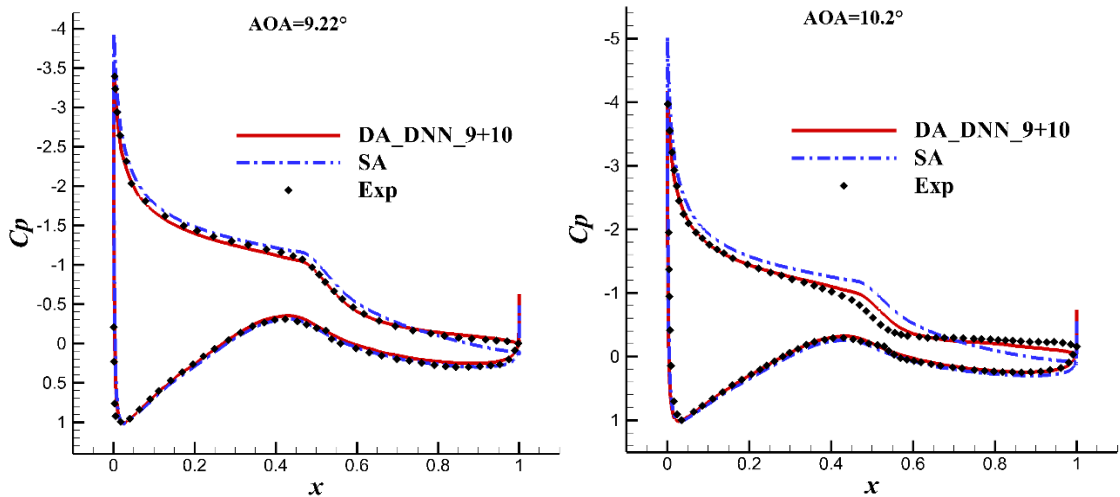

**Fig.9.** The comparisons of pressure coefficients between experimental data and numerical results at 9.22° and 10.20°

In this case, we are more concerned with the generalization capability of the DNN model. The model obtained are used for RANS simulation at other angles of attack. The lift coefficients results and the comparisons are shown in Fig.10. we can see from the results that through the introduction of the experimental data at 9.22° and 10.20°, both the predictions of attached flows at low angles of attack and separated flows at high angles of attack are significantly more accurate compared with traditional SA model. The separation point and the variation tendency of lift coefficients are both in consistent with the experimental data. The relative errors of lift coefficients at high angles of attack are reduced by more than three times. The detailed relative errors of SA model and our DNN model compared with experimental data is shown in Table 2.

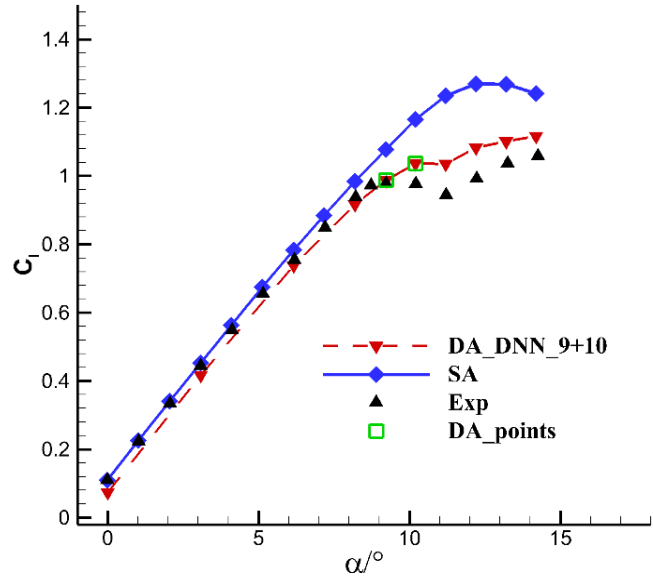

**Fig.10**. The comparisons of lift coefficients versus angle of attack between experimental data and numerical results.

Table 2 The comparisons of lift coefficients and the relative errors

| AOA | Experimental data | Results of SA model | Relative errors of SA | Results of DNN model | Relative errors of DNN |
|---|---|---|---|---|---|
| 3.08° | 0.443 | 0.453 | 2.3% | 0.417 | 5.8% |
| 6.16° | 0.754 | 0.783 | 3.8% | 0.737 | 2.2% |
| 8.20° | 0.932 | 0.972 | 4.3% | 0.917 | 1.6% |
| 9.22° | 0.977 | 1.077 | 10.2% | 0.987 | 1.0% |
| 10.20° | 0.981 | 1.160 | 18.2% | 1.036 | 5.6% |
| 13.20° | 1.039 | 1.282 | 22.2% | 1.100 | 5.9% |
| 14.20° | 1.060 | 1.265 | 19.3% | 1.115 | 5.3% |

Besides lift coefficients, other flow field variables are further compared. The comparisons of pressure coefficients of several typical AOA are shown in Fig.11. The comparison of the velocity field and the streamlines between SA model and the DNN model are shown in Fig.11. We can see from the results that the

$C_p$ results of the DNN model are in consistent with the experimental data especially on the upper surface. The comparisons of velocity distributions and streamlines are shown in Fig. 12. We can see that the separation region simulated with the DNN model is significantly larger than the results of SA model. Furthermore, the eddy viscosity distributions are shown in Fig.13. The main difference of the eddy viscosity distributions between the results of SA model and the DNN model is that the eddy viscosity on the upper surface at the trailing edge and in the wake region is significantly amplified by the DNN model compared with SA model, which helps to predict the separation region accurately.

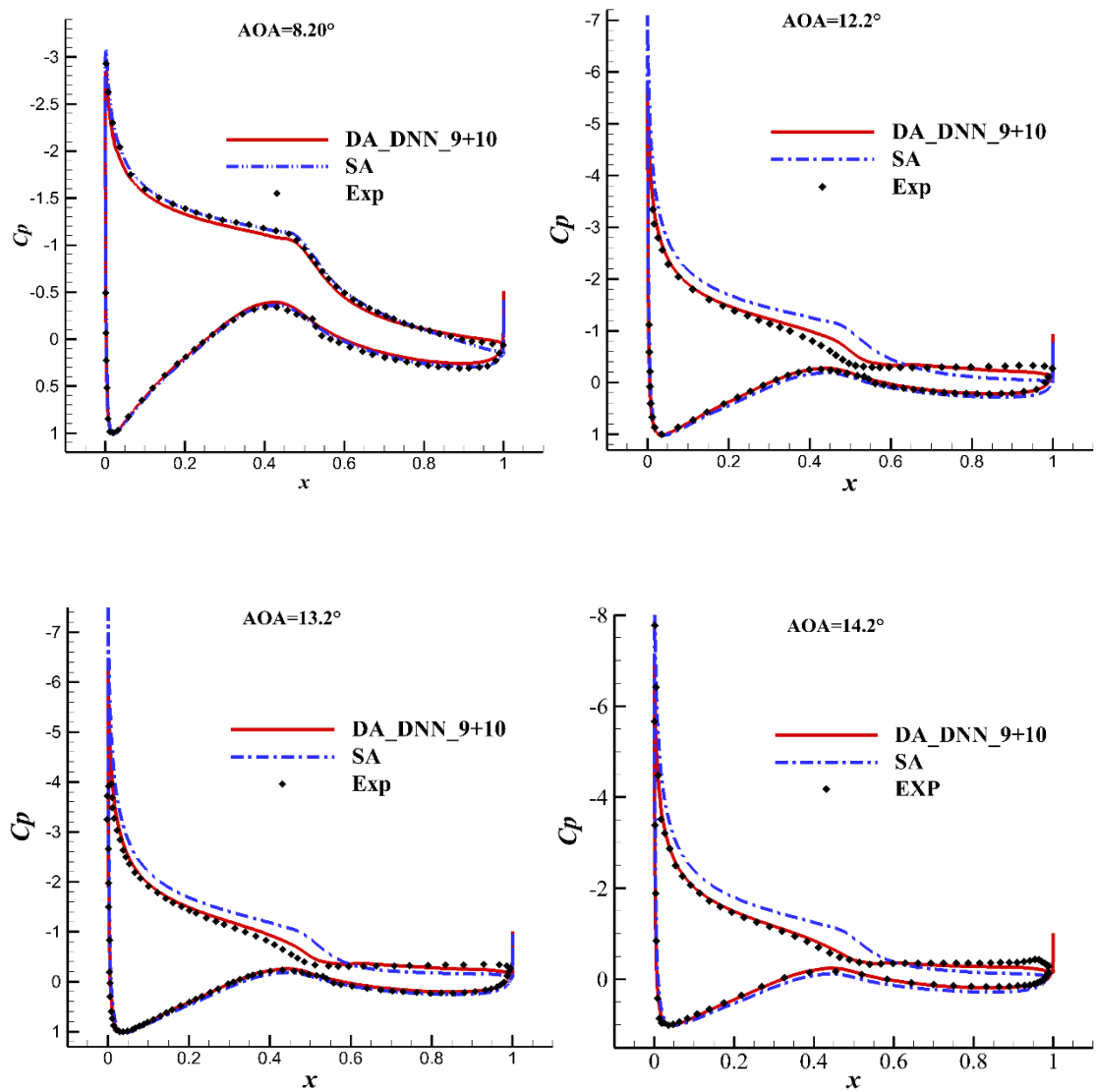

**Fig.11**. The comparison of pressure coefficients between numerical results and experimental data at several extrapolation states

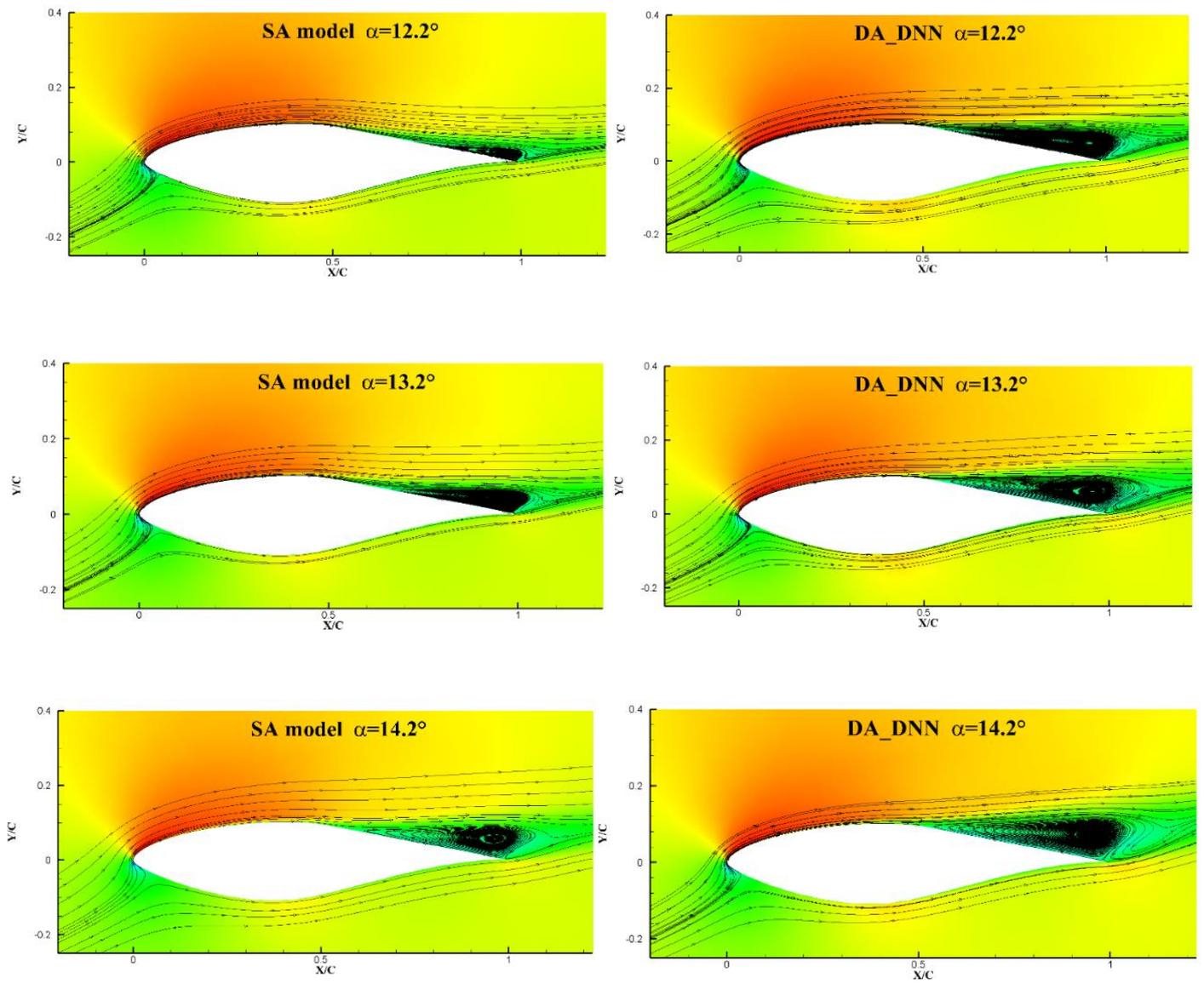

**Fig.12.** The comparisons of velocity distributions and streamlines between SA model and the DNN model

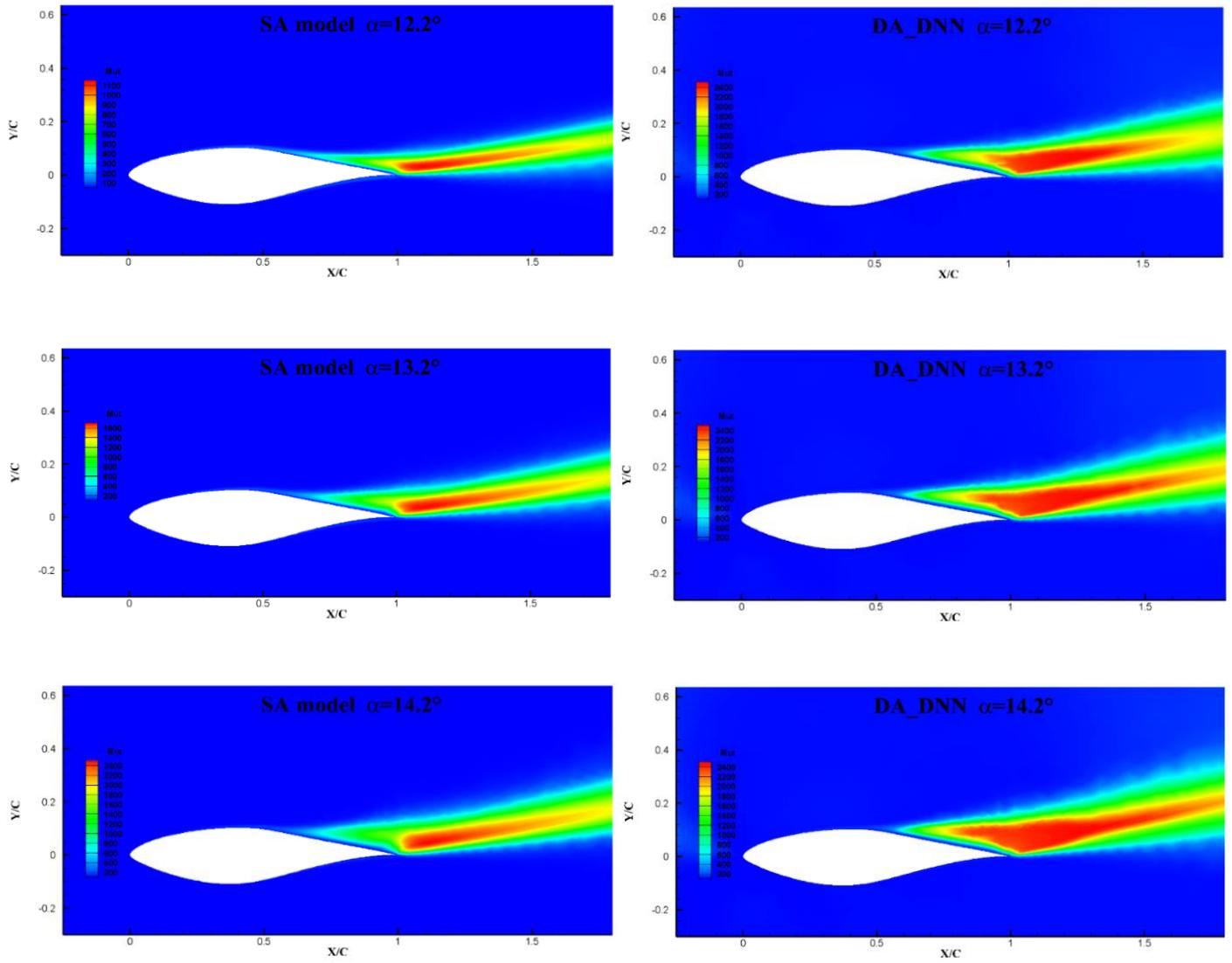

**Fig.13.** The distributions of eddy viscosity for SA model and the DNN model

# 4. Discussion

## 4.1 The computational efficiency of the method

When applied ensemble kalman method for turbulence modeling, the main part of the computational cost is from the solving of RANS equations. As a derivative-free method, tens to hundreds of ensemble members are needed to be propagated in the forward model for the approximation of gradient in tens to hundreds of iterations. Each ensemble member is embedded in RANS solvers in every iteration. The RANS equations are solved until the given convergence tolerance is reached. Despite the advantages of ensemble kalman method in terms of convergence and parallelization, the high computational cost is still a significant limitation of the method.

With the proposed ensemble initialization method based on transfer learning, the information implied in RANS solutions simulated with traditional turbulence models and the differences between RANS solutions at different states are utilized effectively. The ensemble members are initialized with better prior values. Since the difficulty of modeling on noisy and limited experimental data is significantly reduced. Besides, the proposed error covariance setting strategy also helps to improve the efficiency for the solution of the inverse problem. As a result, very few iterations are needed. Actually, only a single iteration is needed in the cases in this paper. The computational cost is significantly reduced by tens of times, even take the preparation of traditional RANS solutions and the pre-training and fine-tune processes into account. Furthermore, in our perspective, too many iterations will be detrimental to the generalization capability of the model, which is similar to the overfitting of neural networks.

## 4.2 The generalization capability of the model

From the two cases in this paper, we can see that the using of high-fidelity data from several representative states as joint constraints is necessary for the improvement of the model's generalization capability. The framework of purely data -driven DNN turbulence model also provide enough trainable parameters and potential space for the solution of the inverse problem compared with the framework of traditional turbulence models. Even so, we should point out that the generalization capability of the DNN model is still limited. There are still difficulties for the DNN model generalizing to more differentiated flow states or geometries. And the present works are based on the eddy viscosity model, which has several limitations itself for its assumptions.

Furthermore, the selection of the observation data is an important factor for such model-consistent training methods, which will significantly influence the performance of the obtained DNN model. Due to the lack of advanced experimental measurements, only pressure coefficients are used for training and comparison in this work. Once we have enough data, diverse data can also be used for training the model together with pressure coefficients without any re-developments of the method or the codes of RANS solvers. Data of different flow conditions including different geometries, different variables such as velocity and friction coefficients as well as data at different spatial positions can be selected to train a turbulence model jointly. While the selection principle of the above aspects and the well-posedness issue of the inverse problem still need further studied.

## 5. Conclusion

In this work, an improved kalman inversion method is proposed with the introduction of transfer learning method and the observation covariance setting strategy. Data assimilation and model training are implemented in a unified way to overcome the inconsistency between training and prediction environments of traditional two-step methods. The method is non-intrusive, and has advantages of convergence and parallelization. Besides, benefits from the improvement measures proposed in this paper, the efficiency are significantly improved, and the properties of the DNN model can be better guaranteed.

The experimental pressure coefficients are used in the two cases of flows around the airfoil at high Reynolds numbers to train the DNN eddy viscosity models. Only the experimental data of very few states is needed with the method to obtain the model with satisfactory generalization capability to both attached and separated flows at different angles of attack. The relative errors of separated flows at high angles of attack are significantly reduced by more than three times. The computational cost is significantly reduced by tens of times compared with existing ensemble kalman methods. Furthermore, the proposed method also has potential advantages from the perspective of data fusion of CFD and experiments.

There are still some limitations and issues of the current method that need further research. In terms of the generalization capability, the selection principle of observation data and the well-posedness issues. Nevertheless, it provides a framework, in which the rich data generated by both CFD and experiments can be used efficiently for the development of a truly generalizable turbulence model.

## Acknowledgements

This work was supported by the National Natural Science Foundation of China National Natural Science Foundation of China (No.92152301, No.12072282).